# THE DUST-GAS FIREBALL AS A SPECIAL FORM OF THE ELECTRIC EROSIVE DISCHARGE AFTERGLOW

S. E. Emelin[a], A. L. Pirozerski[b], N. N. Vassiliev[b]

*Saint-Petersburg State University, Russia*

[a] - *Research Institute of Radiophysics,* <Sergei.Emelin@pobox.spbu.ru>

[b] - *Research Institute of Physics,* <piroz@yandex.ru>

Abstract.

Main results of researches of the electric erosive discharge afterglow plasma, of the fireballs being formed by this plasma and having some particular but deep similarities to the ball lightning are given. The analysis of our results and results of some other authors has allowed to specify the nature of radiation and stability of the shape for a dust-gas chemical fireball, to make series of assumptions concerning the mechanism of these properties. The experimental results and considerations of the given work can be interesting for physics of plasma with different kinds of condensed state, for an ascertainment of the chemiluminescence mechanism of the flame, as well as for a bead lightning understanding and for improvement of ball lightning laboratory analogues.

### 1. Introduction.

In spite of the highest level of many advances of modern science initiated by practical necessities of mankind the problem of the ball lightning remain unsolved. Experience of searching for approaches to this problem, accumulated by now, and appearance of results which seems to be advances in this direction [1–6], make researchers to draw special attention to obtaining and experimental studies of objects which have, except of general resemblance, a deep (but possibly partial) similarity with observed features of the natural, but difficult for studying, phenomenon. Bearing in mind that the electric discharge is the principal cause of the ball lightning origination and optical radiation as its main property, the electric discharge afterglow can be considered as a general phenomenon, in the framework of which physical processes determining nature and totality of the properties of the ball lightning are contained [5,7].

Recently a particular interest is attracted to laboratory analogues of the ball lightning, which are globular formations having high luminosity directly after the discharge termination but then, unfortunately, rapidly decreasing their brightness and losing their form, for example [6]. Their distinctive feature is capability to have a sufficiently sharp quasispherical external boundary of a size characteristic for the average statistical ball lightning. However, they have low energy density $\sim 10^{-1}$ cm$^{-3}$, cannot explode or burn through a foil, have weak macroscopic electric field, exhibit a clear tendency to rise in the atmosphere. In paper [8] we have pointed out the physical nature of this kind of formations as objects appearing due to structure-energy self-organization of a metastable substance on the base of the coulomb and polarization non-ideality of plasma with smallest metal and dielectric particles. The existence of this non-ideality is maintained by catalytic energy-release of decaying chemically active plasma or by electric discharge, moreover the large particles are concentrated near the extern boundary forming "cold envelope" while little ones form "hot core". Subsequent years confirmed validity of this interpretation more and more. In the present paper we discuss a series of experimental data elucidating the nature of principal properties of dust-gas ball lightning.

### 2. Afterglow of a stationary electric discharge in the air flow.

In the air flow directly behind the electric discharge a plasma parameters relaxation zone is

situated – the afterglow area. In paper [9] it was shown by spectral studies that discharge action is accompanied by electrodes erosion and not only results in the gas heating and ionization, but also enriches its composition by chemical components with high formation enthalpy as well as by atoms and molecules in exited states. In spite of relatively high gas temperature ~ 2000 – 3000 K and, correspondingly, short afterglow time ~ 0.1 ms, a high degree of plasma nonequilibrium was obtained. Comparing radiation intensity distributions of the lines Cu I 510,5 nm and the $N_2$ ($C^3\Pi_u$-$B^3\Pi_g$) $2^+$-system band at 337.1 nm along the flow it have been concluded that the afterglow is due to nonequilibrium emission of excited nitrogen molecules. Under conditions of reduced pressure nitrogen flow $NO_2$ adding resulted in bright luminescence at the green-blue spectral range, which was explained by a hemiluminescence reaction between copper atoms from the electrode and nitrogen dioxide molecules with copper oxide formation [10]. Adding of oxygen in the afterglow area strongly increased the radiation intensity of the copper atoms, which was explained by a change of the nitrogen molecules distribution over excited states [11].

### 3. Afterglow of a non-stationary high frequency discharge.

A more prolonged afterglow appears in volume of a molecule gas (including air, nitrogen, oxygen or nitrous oxide), limited by the interelectrode gap, under short-time influence of power high frequency discharge, which the nonequilibrium energy input is typical for. In paper [4] a 10-kW 75 MHz generator was used for that. In the experiment luminous balls of 10-20 cm in diameter arose, their luminosity was sufficiently bright directly after current breaking and last from tenths to two seconds. The afterglow spectra shows solely atomic lines of impurities with excitation energy below 5 eV, first of all lines of electrode's refractory materials, and, when oxygen was present, also bands of their oxides. The afterglow spectrum was constant, only its intensity was changing. Gas temperature inside the nonequilibrium luminous formation $T_g$ ~ 2000–2500 K was found by heating of a tungsten wire 0.075 mm in diameter placed inside the formation. It is the author's opinion that measured radiant energy was supplied by excited molecules of nitrogen and oxygen with relative concentration about tenth of percent. By our opinion, for understanding energetic nature of these object it is important the fact that when using nitrous oxide, whose molecule has an excess energy ~1 eV, the afterglow time increased by the factor ~4 and the luminous area diameter rose up to 0.5 m. Note that under normal conditions this volume of $N_2O$ contains 0.25 MJ of energy.

### 4. Erosion capillary discharge afterglow.

This discharge, presented in paper [2], was performed in a cylindrical hole ~ 1 mm in diameter at a polymer plate or tube. One end of the holed was hermetically closed by a metallic or carbon cathode, and another one was supplied by a ring copper anode. A capacity tank through an inductance discharged on to the capillary. Discharge products formed a long rapid luminous jet with some interesting properties [3]. The gas temperature being low (below 1000 K), the jet radiation spectrum out the discharge zone shows atomic lines C I, Cu I, Ca I, Na I and molecular bands $C_2$ and CN. In the zone near to the discharger high broadened lines of atomic hydrogen with excitation energy exceeding 12 eV were distinguished. A weak continuous spectrum of hot carbon particles was present. By infrared radiation presence of the carbohydrates $CH_2$, $C_2H_4$, $C_3H_8$, $CH_3OH$ was found. The spectra time dynamics study at the range 420–540 nm in the jet kernel local zone at the distance 40 mm from the discharger revealed the presence of bursts with duration about 10 μs. Their radiation besides of a continuous spectrum had very strongly broadened lines of atomic hydrogen 486.1 nm and of carbon 538.0 nm. Nature of the bursts was presumably explained explosions or by the pass of high-temperature clots through the plasma. To obtain an autonomous formation the front part of the plasma jet, passed through a thread wetted by the saturated solution of NaCl, was detached by a high-speed cocurrent air flow. The autonomous object lived ~ 3 ms. Its spectrum shows the lines Na I, Ca I, Ba I, red bands of CN, a weak CaCl band – an indication of the free chlorine presence, and, which is most important, the lines of atomic hydrogen 656.22 nm, that evidences the presence of a high energy and hydrogen concentration.

The authors of the present article have discovered by electric and microwaves measurements a unipolar generation of low-frequency oscillations of plasma parameters and generator mode near the current breaking [12]. The studies of dynamics of brightness clots (transfer waves) along a relatively slow jet of the low-current discharge by means of a system of optic and electric probes [13,14] show that the clots cause anisotropy of the plasma conductivity, can have a speed several times higher than that of the jet, can gather in the front part of the jet forming an excitation wave, a kind of leader feed from the side of the plasmogenerator and capable to a short-time autonomous existence after the current breaking or after detachment from the jet. Nature of this structure-energy state was conceivably characterized as a metastable substance on the base of a condensate of high excited atoms and molecules – the Rydberg matter [15].

### 5. Study of the non-stationary discharge from a 380V power mains.

#### *5.1. Experimental setup for obtaining and studies of the fireball.*

To obtain and to study the dust-gas fireball we used a non-stationary electric discharge from the laboratory 380V power mains [7]. The discharge was performed by a short-time contacting and further moving apart, up to breaking-off of the current, of electrodes, connecting to the mains via a series-resonant circuit. The circuit capacitor was 200 μF x 5 kV, the ohmic resistance of the coil being 6 Ohm. Carbon rods of diameter 8 mm, steel, copper or aluminum wires 1.2–2 mm in diameter depending on the metal melting temperature were the electrodes materials.

To study of the physical nature of the chemical dust-gas fireball we carried out photo and video recording of the luminous objects by digital camcorders using optical filters, relative general and spectral radiation intensity were measured by a photodiode detector and by a photoelectron multiplier, the radiation intensity and current kinetics were studied by two-channel oscillography.
For studying the discharge and the afterglow plasma spectrum dynamics an original spectrograph consisting of two separate modules was used. The first one was a support with two metal sheets 0.5 m x 0.5 m installed vertically with an adjustable gap between them. They formed the spectrograph slit behind which the discharge was realized and a He-Hg lamp with a reflector was installed as a reference source. The second module represented a metal plate – a round table – installed on a telescopic tripod. The table had a possibility to turn at any angle around of a vertical axis and to be inclined at the angle - 45º - + 45º around of one of horizontal axes and to be fixed. On the table the diffraction grating with 2400 or 1200 grooves per millimeter was established with a possibility of rotation around of the vertical axis laying in the middle of a reflection plane of the grating. Located on the table horizontally the bar could turn around of a vertical axis of a grating with an ability of a fixation. On the bar the camcorder or the camera with possibility of rotation in a small angle and of shift along the bar for moving off or nearer to the grating with an ability of fixation was installed. This construction enabled to register the spectral distribution along the vertical direction at the spectral range 200 –700 nm for the 2400 gr/mm grating and 400 –1400 nm for 1200 gr/mm one. The camcorder Sony DCR-TRV11E allowed to take spectra at the range 380 – 650 nm with color displaying and at infrared range (up to 1100 mm) under the "night shot" regime, the frame field rate being 50 field per second and the maximal spectral resolution being ~0.5 A.

#### *5.2. Experimental study.*

*5.2.1. General features of the discharge.* This discharge consists of several stages. Contacting of the electrodes results in the appearance of a current which sets in during a few periods attains the magnitude ~ 60 A (effective) (see Fig.1), as well as in heating of the electrodes contact area.

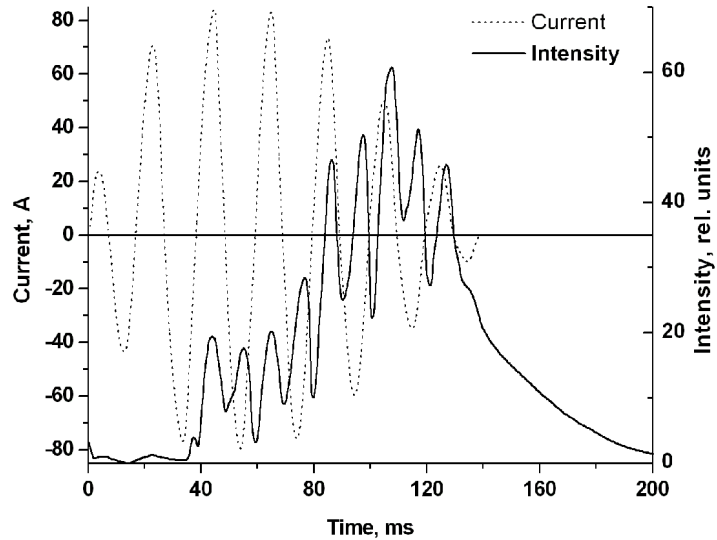

Fig.1. Current and intensity kinetics.

The subsequent separation of the electrodes leads to arising of a lengthening floating up non-contracted arc accompanied with an increased erosion of the electrode material. At the upper part of the arc discharge products accumulate forming a bright luminous plasma volume, in which damping discharge current is distributed (Fig.2).

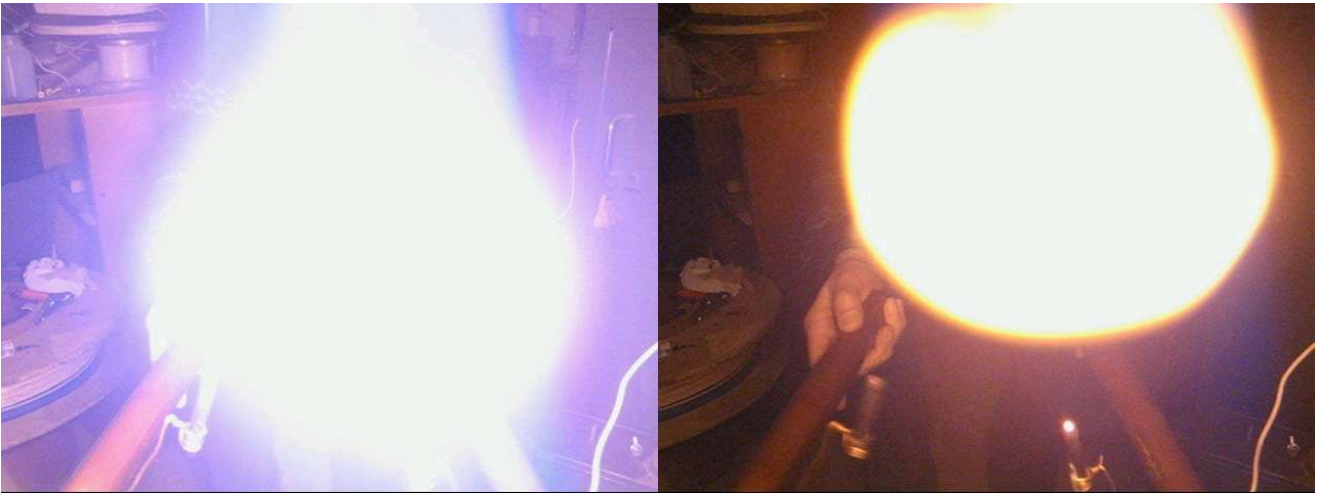

Fig.2. Volumetric discharge with carbon electrodes accompanied with combustion and discharge's afterglow.

This volume of size ~10 – 20 cm tends to take a spherical shape, but it can be easily disturbed by extrinsic air streams, non-optimal orientation of the erosive jets or by a mismatch of the transient plasmodynamic process with the technique of moving apart of the electrodes. When the current magnitude becomes sufficiently low, the current breaking occurs at its transition through one of the zero points, then the afterglow stage begins. The ozone smell is felt near the discharge.

*5.2.2. General features of fireballs.* During the afterglow stage the fireball brightness decreases, in some cases its color changes strongly (Al, Cu). The fireball shape relaxation can occur by different possible scenario. If at first short period, when the surface tension is maximal, the inner gas dynamics is not relaxed yet with formation of single core then next the object splits at several parts, sometimes flying away in different directions, or takes the form of departing jet. The object with inner toroidal vortex has a spherical envelope during the first 50–100 ms, which then transforms to a toroid (Fig. 3).

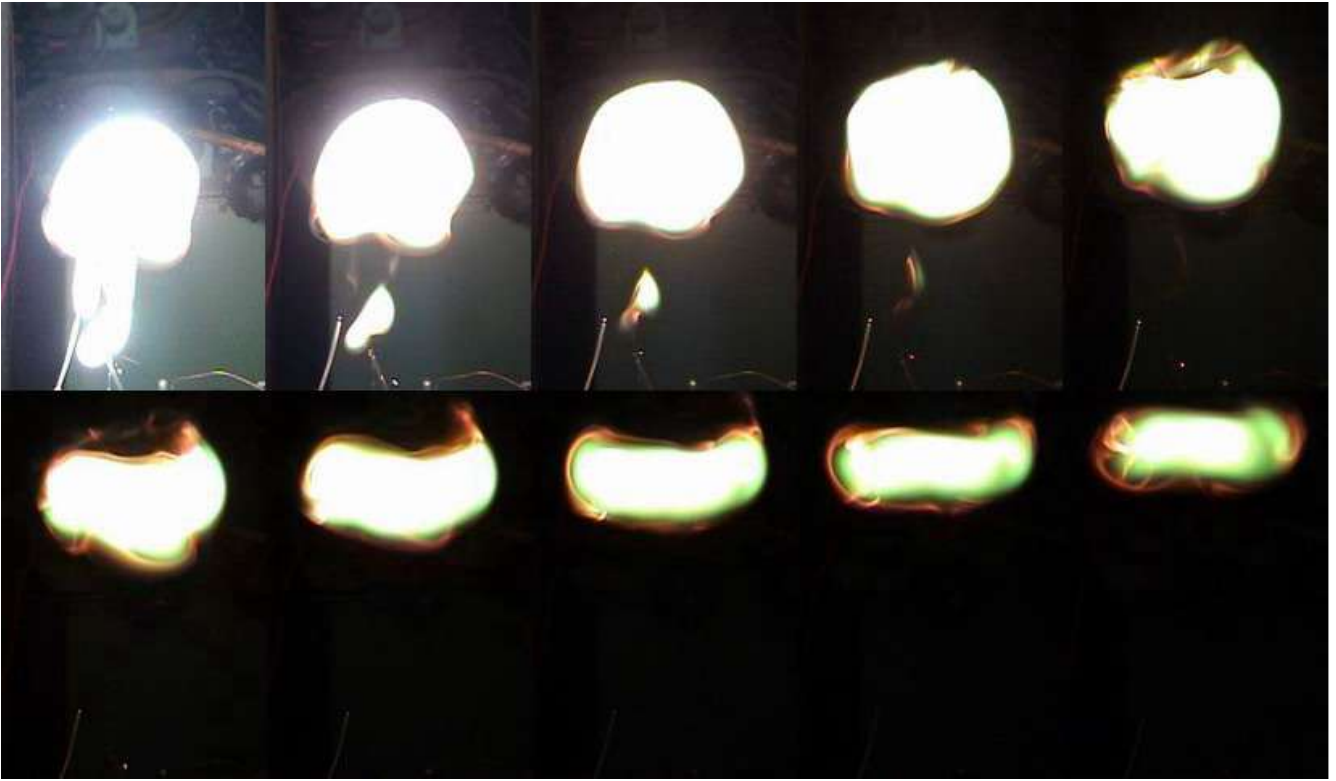

Fig. 3. Transformation of a fireball to a toroid at the discharge with copper electrodes.

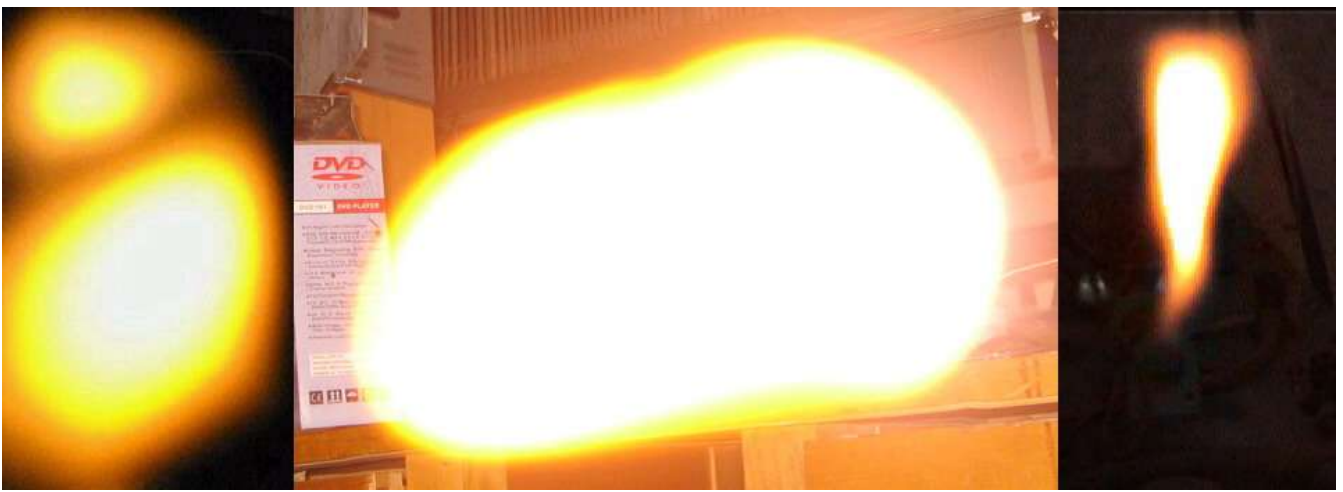

Fig. 4. Different forms of the afterglow of the discharge with carbon electrodes.

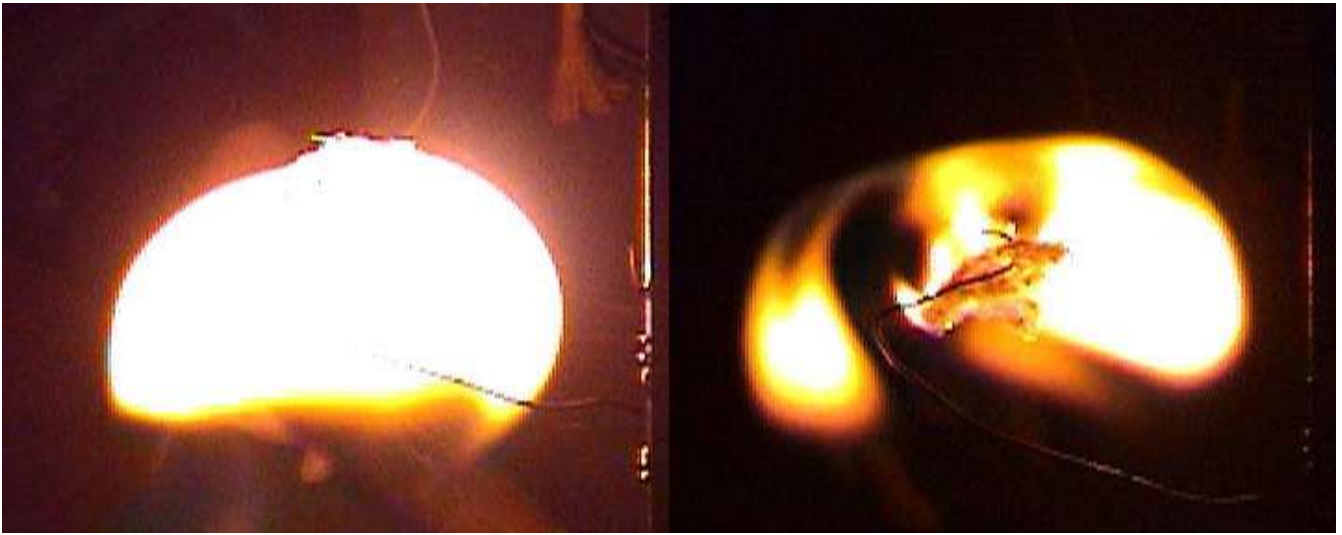

Fig. 5. Cotton wool firing.

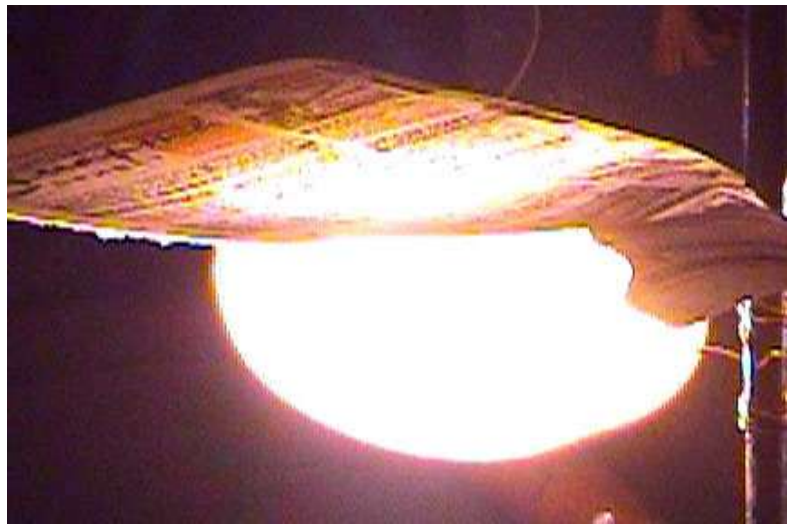

Fig. 6. Heat pulse is insufficient to set fire to a sheet of newsprint.

Brightness of a "calm" object on the basis of the carbon grows to its center. The object relaxes with decreasing of the size. Its ascent can be accompanied by the shape changing to elliptic, tear, etc. (Fig. 4). Objects, containing a minimal amount of the dust substance and a large volume of gas seized by the discharge, consist only of an envelope, sometimes non-closed at the bottom, similarly to [6]. The objects are capable to set fire on cotton wool, but only at the beginning of the relaxation stage (Fig. 5). Unlikely to the discharge plasma, they can not set fire on a sheet of newsprint (Fig. 6) or burn through aluminum foil. During firsts tens of milliseconds the objects have the thixotropic property, i.e. the capability to pass through themselves little objects without self-destroying (Fig. 7). A similar discharge was carried out in a nitrogen filled chamber but the discharge conditions optimization for generation of spherical-shaped objects was not performed.

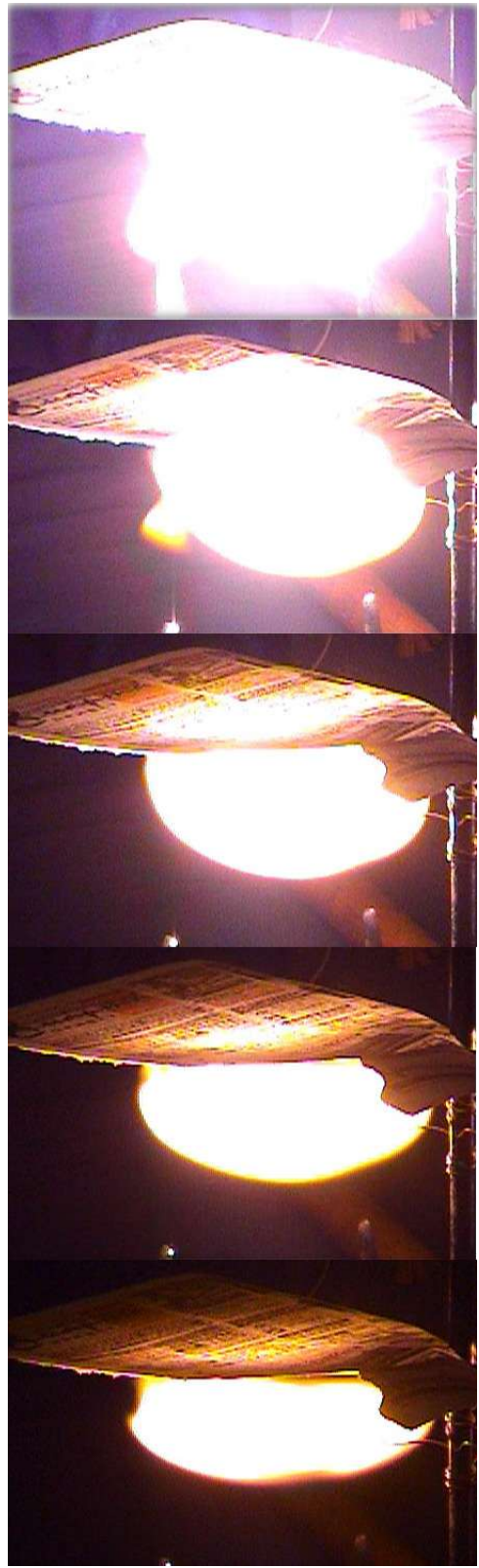

Fig. 7. Surface tension and thixotropy relaxation.

*4.2.3. Study of the discharge and fireballs radiation.* The discharge radiation has two components. The first one changes near synchronously with the absolute value of the current (the delay is about 1–3 ms). The second one varies only little during one half-period of the current (Fig. 1). This component increases smoothly during the pulse of the current and causes the afterglow existence. On the total radiation intensity oscillogram three parts can be discriminated which differ one from another by type of the time dependence. This is revealed itself especially strongly when carbon electrodes are used and is connected with an influence of the dust-gas dynamics on the burning rate.

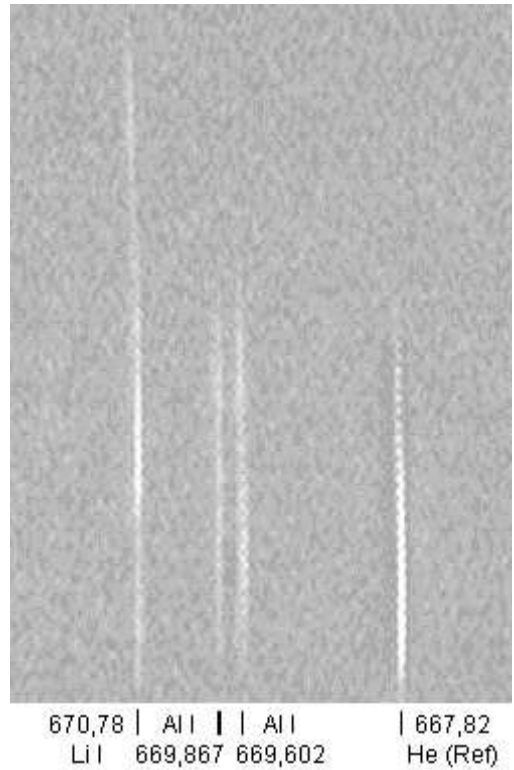

Fig. 8. A part of spectrum of the discharge with aluminum electrodes.

The presence of short-living and long-living components was found at all spectral lines, between them there were a difference in the ratio of these components and in afterglow duration. For example, with aluminum electrodes the bright lines Al I 669,602 nm and Al I 669,867 nm ($E_{ex}$ = 4.99 eV) (Fig. 8) appear only in the near to electrode regions of the arc at the period of the current maximum, but these lines are weak or absent at the distributed current area and at the afterglow. At the beginning the afterglow is due to the brightest bands of AlO (dissociation energy $E_{dis}$ = 5.24 eV) (see Fig. 9), but then it is due only to alkali metal lines, mainly to Li I 670,776 nm and 670,791 nm ($E_{ex}$ = 1.85 eV), and Na I 589.0 nm and 589.6 nm ($E_{ex}$ = 2.1.eV), that causes the change of the object color from blue to yellow-orange.

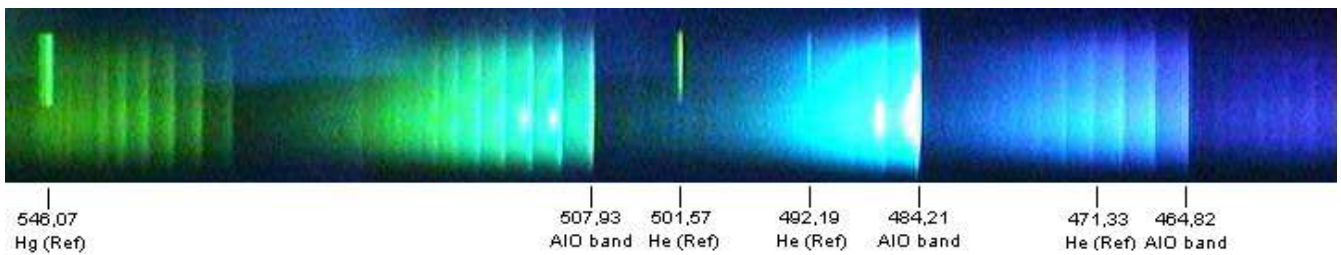

Fig. 9. The afterglow spectrum at the green-blue range, aluminum electrodes.

In the case of copper electrodes, along with bright bands CuO ($E_{dis}$ = 2.77 eV) with cants 614.68 nm and 616.15 nm, and CaO ($E_{dis}$ = 4.11 eV) with cants 625.85 nm and 627.8 nm, at the afterglow beginning the atomic lines Cu I 402.27 nm and 406.27 nm ($E_{ex}$ = 6.87 eV), Cu I 515,32 nm and 521.82 nm ($E_{ex}$ = 6.19 eV) existed, and Ca I 422.67 nm ($E_{ex}$ = 2.93 eV) were observed a longer time. The lines with a higher excitation energy Cu I 510.55 nm and 570.02 nm ($E_{ex}$ = 3.82 eV), 578.21 nm ($E_{ex}$ = 3.79 eV) as well as the Na I line existed for a more long time. Similarly to [10] a bright system of bands was persistently observed at the green-blue spectral range, which is seemingly due to $Cu_2O$ (Fig. 10).

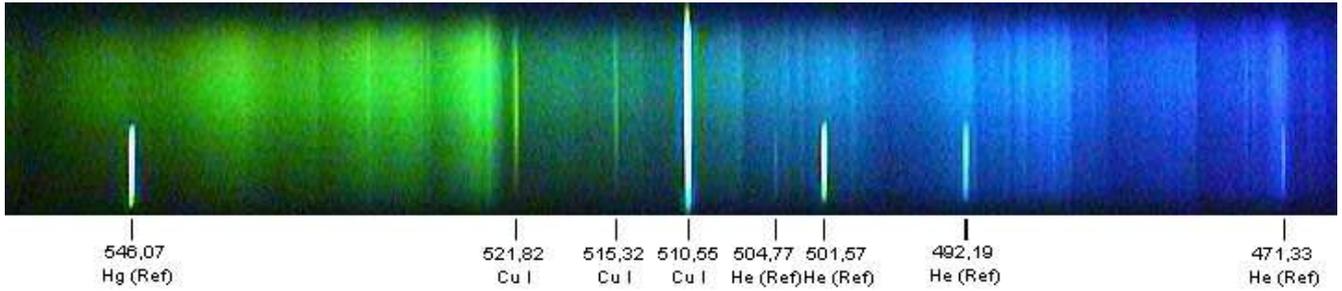

Fig. 10. The afterglow spectrum at the green-blue range, copper electrodes.

When carbon electrodes were used, beside of molecular bands at the afterglow very bright atomic lines were present – only lines of impurities Fe I, Cu I, Ca I, Ba I, Cs I, Rb I, Li I, Na I, K I, and the last four have longest lifetime.

In all cases for objects with the core atomic lines with highest excitation energies were concentrated in the core. With copper electrodes videoshooting through optical filter at the range 1 – 1.1 μm gives images on which only the object envelope was visible.

## 6. Discussion.

Comparing our analogues with the true BL, properties of which follow from evidences of eyewitnesses, we should state that they can be interesting only on three main aspects: the nature of energy, radiation and shape stability of some possible BL types. Character of radiation of our objects appeared to be purely luminescent, and its spectrum was qualitatively similar to [4]. The radiation is mainly presented by a molecular spectrum of oxides and an atomic spectrum of metals. Presence of two-atomic oxides indicates that they forms from the atoms arising at strong heating of metal in an electrode spot and current filaments and as a result of oxygen molecules dissociation under action of the UV radiation. Though absence of the excited atoms does not mean literally absence of the atoms, nevertheless disappearance of lines Al I long before the ending of the discharge forces to think that metal vapor is completely burned down even before the beginning of afterglow (in contrast to [5] where for prevention of that the erosive ejection formed a toroidal vortex). Therefore the luminescence of oxides and atoms in the afterglow indirectly points to the presence of a source of excitation and even of decomposition of molecules in this period.

Note that in contrast to the aluminium (dissociation energy AlO $E_{dis}$ = 5.24 eV), at the afterglow plasma lithium atoms are present ($E_{dis}$ (LiO) = 3.55 eV), with other electrodes – other alkaline metals, copper, calcium, iron. It means that with a significant probability two-atomic oxides are capable to take from the pointed source energy E being some less than

$$E < E_{max} < E_{dis}(AlO) + E_{ex}(Al) = 5.24 \text{ eV} + 4.99 \text{ eV} = 10.23 \text{ eV},$$

but even at the end of the afterglow surpassing

$$E > E_{min} > E_{dis}(LiO) + E_{ex}(Li) = 3.55 \text{ eV} + 1.85 \text{ eV} = 5.4 \text{ eV}.$$

In the case of copper electrodes:

$$E_{max} > E_{dis}(CaO) + E_{ex}(Ca) = 4.11 \text{ eV} + 2.93 \text{ eV} = 7.14 \text{ eV},$$

and

$$E_{max} > E_{dis}(CuO) + E_{ex}(Cu) = 2.77 \text{ eV} + 6.19 \text{ eV} = 8.96 \text{ eV}.$$

Table 1 shows two-atomic oxides dissociation energy $E_{dis}$, minimal atomic line excitation energy $E_{ex}$ of the corresponding element and their sum $E_{min}$ for series of metals.

Table 1.

| Oxid(radical) | $E_{dis}$, eV | $E_{ex}$ eV | $E_{min}$, eV | Oxid(radical) | $E_{dis}$, eV | $E_{ex}$ eV | $E_{min}$, eV |
|---|---|---|---|---|---|---|---|
| RbO | 2,63 | 1,56 | 4,19 | FeO | 4,21 | 3,21 | 7,42 |
| CsO | 3,07 | 1,39 | 4,46 | AgO | 2,47 | 3,66 | 6,13 |
| KO | 2,9 | 1,6 | 4,5 | AuO | 2,47 | 4,63 | 7,1 |
| NaO | 2,64 | 2,1 | 4,74 | PtO | 3,84 | 4,04 | 7,88 |
| LiO | 3,55 | 1,85 | 5,4 | BaO | 5,81 | 2,24 | 8,05 |
| CuO | 2,77 | 3,79 | 6,56 | AlO | 5,24 | 3,14 | 8,38 |
| CaO | 4,11 | 2,93 | 7,04 | ZnO | 2,85 | 5,8 | 8,65 |
| MgO | 3,73 | 2,71 | 6,44 | SiO | 8,29 | 4,13 | 12,42 |

Considering the energy nature of the mentioned source we pay attention to the small time of the afterglow caused by excited molecules of nitrogen ∼ 100 μs [9], a little bit greater lifetime of the atomized gas in air [3], the increase of lifetime and the size of a fireball in several times when using nitrogen protoxid [4], the appearance of nitrogen oxides in the air stream under action of the discharge [9], a strong ozone smell. Taking into account that our objects hardly set fire even to cotton wool and rather slowly ascent in the atmosphere, we consider that their gas temperature does not exceed 300-400°C already in the beginning of afterglow.

This is explained by small energy inputs ~ 1 kJ and by the distributed character of the discharge accompanied with combustion [20] and also exerting a significant chemical influence onto the air and the electrodes substance. The temperature is high only in current-dust filaments where some concentration of nitrogen oxide are created and hardened when mixing up with the surrounding cold air [16]. Radiation of the discharge creates atomic oxygen which is quickly being spent for ozone formation. Interaction of hot dust carbon particles with nitrogen results in creation of cyanogen which, at temperature existing in the distributed discharge, can form amorphous polymer paracyanogen capable to burn in pure ozone by a red flame with very high temperature. Therefore we think that the action of the discharge consists in creation in the air of some concentration of chemically active substances (NO, $O_3$, CN etc.) and fine aerosol of oxides, nitrides, the air heating being small.

Note that the chemical energy of any possible in this case elementary reaction does not surpass 3 eV, it is much less than even minimal energy necessary for appearing of an atomic line ($E_{ex}$(Rb) = 4.19 eV). This directly indicates a collective character of energy accumulation and of its transfer to the luminescence channel. As the fireball radiation is mainly due to free excited atoms and molecules, their occurrence can be explained by the emission of them [17] by hot or electron-excited particles of the aerosol. Thus the afterglow is caused by burning of chemically active gaseous and aerosol (aerogel) substance, during of which the fine aerosol particles eject by process of thermoemission or cooperative chemoemission the excited atoms and molecules creating the luminescence.

If we attribute this effect to thermoemission then starting from the relative intensity of the lines Cu I 515.32 nm, 521.82 nm and 510.55 nm and taking into account dissociation energy of the oxides we find that the electronic temperature of aerosol particles $T_e$ falls during the afterglow period from ∼ 0.8 to ∼ 0.4 eV or less. High nonequilibrium of such "plasma" can be explained by the nano-size of the particles: having a large surface to volume ratio they quickly give away the vibrational energy to surrounding cold gas and release the excessive electronic excitation by means of the emission of excited atoms, molecules, ions, electrons or by explosion [3]. For larger particles of an aerosol (impurity semiconductors, electrets) the effect can be explained by increased stability of their metastable electron-excited states generated due to the chemoexcitation, by their threadlike shape or by their ability to luminesce without ejection of the excited particles, that is more preferable for FB lifetime increasing [18].

Some spectral lines observed at the afterglow beginning have the value of work function plus excitation energy considerably exceeding the one for ions, for example, alkaline metals. Note that such ions should be accelerated (be slowed down) by an electric field of a positively (negatively) charged aerosol particle which throws them out, characteristic potential of the particle being of order ~ eV. In our opinion, ions recombination under non-ideal plasma conditions can be accompanied by formation of clusters of the Rigberg substance [15]. Their disintegration to atoms should be accompanied with the radiation from high excited states which was observed by us in the beginning of the afterglow.

Being a property of disperse systems, the thixotropy, shown by our objects and contributing to the stability of their shape, indicates the existence of a dust plasma condensate at least in the peripheral part – the envelope. Most likely it consists of a polymeric aerosol [19, 20] on the basis of nitride-oxide-impurity electrified particles of two types: donor and acceptor. The thixotropy of the dust-plasma condensate explains the ability of BL to take various relatively steady shapes and imposes the corresponding specificity on its hydrodynamics.

The Fig. 1 shows the presence of two channels of the luminescence excitation: by short-living (electron-excited molecules and atomic radicals) and long-living (oxides of nitrogen, ozone, burning aerosol) active particles. Since long afterglow is caused exclusively by the last mechanism, it can not be consider as correct to identify with BL the objects before the relaxation stage following the termination of the energy input from an electric source, except when this source is plasma or nuclear one [21].

## *7. Conclusion.*

The analysis of the results of the researches of the erosive electric discharge afterglow plasma from a power mains 380 V, of formed by this plasma the autonomous dust-gas fireballs having particular but deep similarities with BL in respect of appearance, luminosity and relative stability of the shape, and also analysis of results of some other authors has led us to the following conclusions:
– the nature of the radiation of a fireball considered as special form of a long afterglow – a bright luminescence of oxides and atoms of metals,
– energy source of FB is chemically active substance created by the discharge from the air and the erosive emission products – nitrogen oxides, ozone, nitrides, vapor and aerosol of metals, organics,
– the luminescence has a chemical nature (chemiluminescence) and besides of elementary chemiluminescence reactions is mediated by the energy accumulation by particles of an aerosol in the form of vibrational or electronic excitation in consequence of their burning and catalitic reactions to their surfaces. One of channels of excessive electronic excitation release is the ejection of electrons, ions and of the luminescence activators – excited molecules and atoms (mainly light-ionaizable impurities – alkaline metals atoms) – by the dust particles,
– the fireball structure can be of three types. At small quantity of a dust and relatively large volume of the gas seized by the discharge the dust gathers on border of object, forming only an envelope on the basis of non-ideal plasma. Objects of the second type consist of an envelope and an internal toroidal vortex with a combustible dust. During the relaxation such object is capable to change the shape from spherical to toroidal. Fireballs of the third type do not contain internal dynamic perturbations and consist entirely of low dense complex plasma condensate. At a narrow range of relaxing parameters such objects have property of thixotropy and a special hydrodynamics [21].